# Blockchain: Data Malls, Coin Economies and Keyless Payments

Zura Kakushadze[§¶1] and Ronald P. Russo, Jr.[†2]

[§] *Quantigic® Solutions LLC,[3] 1127 High Ridge Road, #135, Stamford, CT 06905*

[¶] *Free University of Tbilisi, Business School & School of Physics*
*240, David Agmashenebeli Alley, Tbilisi, 0159, Georgia*

[†] *GLX – Global Listing Exchange*
*340 Royal Poinciana Way, Suite 317-335, Palm Beach, FL 33480*

January 20, 2018

## Abstract

We discuss several uses of blockchain (and, more generally, distributed ledger) technologies outside of cryptocurrencies with a pragmatic view. We mostly focus on three areas: the role of coin economies for what we refer to as data malls (specialized data marketplaces); data provenance (a historical record of data and its origins); and what we term keyless payments (made without having to know other users' cryptographic keys). We also discuss voting and other areas, and give a sizable list of academic and nonacademic references.

---

## 1. Introduction

Blockchain is a ledger for keeping a record of all transactions. It is a time-sequential chain of blocks containing transaction records. Blocks are linked using cryptography and time-stamping (see Figure 1). The data in a given block cannot be altered retroactively without altering all subsequent blocks,[4] which makes blockchain resistant to modification of data by design. For historical reasons [Nakamoto, 2008],[5] transactions in blockchain are usually thought of as relating to peer-to-peer (P2P) payments with blockchain as a distributed ledger maintained by a P2P network that itself validates new blocks. However, blockchain is not synonymous with payments or decentralization. A priori, blockchain can record other types of transactions. Also, blockchain need not be maintained by a P2P network but, instead, by a centralized authority.

In decentralized blockchain-based cryptocurrencies such as Bitcoin (BTC) and Ethereum (ETH), verification of transactions by a P2P network is achieved via mining, a process whereby GPUs/CPUs solve computational problems. As part of this process, miners (among others) maintain copies of blockchain. Miners receive fees in respective cryptocurrencies. Mining is computationally and energy costly and has been argued to be inefficient and possibly even unsustainable as blockchain grows (see, e.g., [Bariviera, 2017], [Kostakis and Giotitsas, 2014], [Nadarajah and Chu, 2017], [Urquhart, 2016],  [Vranken, 2017]). In contrast, anticipated centralized government-issued cryptocurrencies (e.g., CryptoRuble) do not require mining and in some sense are more "efficient" as transactions are verified by a centralized authority (e.g., a central bank) [Kakushadze and Liew, 2017].[6] In government-issued cryptocurrencies one of the main allures of decentralized cryptocurrencies – anonymity of transactions – is gone. However, employing blockchain (vs. traditional bank ledgers) can help prevent/reduce fraud, errors, etc.

The decentralized v. centralized cryptocurrency example highlights that, as with most things in life, whether using blockchain for solving a particular problem adds value depends on practical considerations and the answer can vary from instance to instance. In this article, we discuss some potential applications of blockchain outside of cryptocurrencies from a pragmatic viewpoint. We mostly focus on three areas: the role of coin economies for data malls; data provenance; and what we term keyless payments. We also touch upon applications to voting.

---

[4] Thus, when blockchain is maintained by a large network as a distributed ledger (see below) continually updated on numerous systems simultaneously, a nefarious alteration would require collusion of the network majority.

[5] For a much earlier work on an anonymous cryptographic electronic money or electronic cash system known as "ecash", see [Chaum, 1983, 1985], which served as a basis for Chaum's electronic money company DigiCash, Inc. founded in 1989. In 1998 DigiCash filed for Chapter 11 bankruptcy and was sold for assets in 2002 [Pitta, 1999].

[6] Also see references therein. For subsequent related articles, see, e.g., [Bershidsky, 2017], [Holmes, 2017].

## 2. Data Malls

Data marketplaces (or data markets) have been discussed in, e.g., [Balazinska, Howe and Suciu, 2011], [Koutris *et al*, 2012], [Koutroumpis, Leiponen and Thomas, 2017], [Li *et al*, 2017], [Liu *et al*, 2017], [Maruyama, Okanohara and Hido, 2013], [Morrison *et al*, 2011], [Muschalle *et al*, 2012], [Nakamura and Teramoto, 2015], [Perera *et al*, 2014a,b], [Schomm, Stahl and Vossen, 2013], [Smith, Ofe and Sandberg, 2016], [Stahl, Löser and Vossen, 2015], [Stahl *et al*, 2016], [Yu and Zhang, 2017], [Zuiderwijk *et al*, 2014]. The idea of a data marketplace in the context of the Internet of Things (IoT) has recently gained traction with the IOTA Data Marketplace (IOTADM) [Schiener, 2017], [Sønstebø, 2017] launched by the IOTA Foundation [Ponciano, 2017], which circulates its own coin MIOTA.[7] The technology underlying MIOTA and IOTADM is the so-called "Tangle", which is a distributed ledger.[8] IOTADM is an open and decentralized data lake (see, e.g., [IBD, 2017], [Ramakrishnan *et al*, 2017]) accessible to any paying party. Its scope is broad.

One of the challenges of broad data marketplaces is precisely their breadth of scope. E.g., it would be difficult to convince a mainstream capital markets professional to purchase stock market pricing data from a data marketplace that is a hodgepodge of vastly disparate data types (hypothetically) ranging from measurements made by a sensor mounted on a weather balloon hovering somewhere above Antarctica to some quantitative properties of bacteria found in the guts of livestock in remote regions of Tibet and everything in between… Furthermore, such broad data marketplaces based on data lakes would normally store data in its raw, unprocessed form,[9] so the burden of making sense of such data (cleaning it, formatting it, etc.) falls on the end-user. For some applications (and industries) this may work just fine. But in other cases, e.g., in capital markets, finance, quantitative trading, etc., end-users desire more of a plug-and-play type of data, which they can utilize without expending prohibitive effort.

---

[7] As of January 6, 2018, MIOTA was ranked as #10 by market capitalization among all cryptocurrencies. See Table 1.
[8] However, it is a form of a direct acyclic graph (see, e.g., [Bang-Jensen and Gutin, 2009]) as opposed to blockchain.
[9] A (perhaps very) crude analogy would be an encyclopedia that is not alphabetical; or a search engine without some form of built-in backend intelligence for automatically determining what the end-user is searching for, which at times can be quite unstructured, e.g., in the form of a convoluted or ill-posed question, jumbled keywords, etc.

More specialized outlets we refer to as Data Malls would appear to be more suitable for this purpose.[10] For example, a Data Mall can act as a decentralized marketplace for various types of financial data (and datafeeds), but not as disparate as in the case of a general, hodgepodge type of a data marketplace. Furthermore, each data provider can normalize and format its data according to the adopted industry practices, so that the data is marketable and easy to use.[11] In some cases, it may be beneficial to have normalized data descriptions (or "abstracts") for ease of searching, classification, etc., with (automated) centralized verification. Such data need not be hosted in a data lake. Instead, it can be hosted on private nodes that the providers control. They choose what content to make public and what content to keep private. Decentralization then is not in dumping data using some kind of a distributed ledger – in most cases this is impracticable for a host of reasons, including the fact that data can be huge in size, not to mention issues that can arise with data provenance, disparate formatting, etc. Instead, decentralization is in creating a network that allows users to buy and sell data from each other directly, without a middleman. This is precisely where Coin Economies can play a pivotal role.

## 3. Coin Economies

So, in some sense, Data Malls provide a middle ground, a hybrid model, between the status quo and the radical idea of completely unstructured, hodgepodge data marketplaces based on data lakes. Blockchain (and similar distributed ledger) technologies can provide decentralized user networks connecting data sellers and buyers. A Coin Economy is a simple way of achieving this.

Thus, blockchain-based Ethereum smart contract technology (see, e.g., [Buterin, 2014], [Christidis and Devetsiokiotis, 2016], [Cuccuru, 2017], [Fairfield, 2014], [Koulu, 2016], [Levy, 2017], [Marvin, 2016], [Mik, 2017], [Omohundro, 2014], [O'Shields, 2017], [Piasecki, 2016], [Raskin, 2017], [Savelyev, 2017], [Sillaber and Waltl, 2017], [Szabo, 1996, 1997], [Werbach and Cornell, 2017], [Wood, 2014], [Xie, 2017a,b]) has been used to create (P2P) networks of users with common interests. Here are 2 examples [Xie, 2017b]. Augur is a decentralized prediction market with an Ethereum token called Reputation (REP). REP is used to reward decentralized reporters (which are all REP holders) of the prediction market event outcomes. Those who report outcomes incorrectly are penalized by taking away some of their REP holdings. Majority of REP holders have incentive to report outcomes correctly: otherwise REP would lose value. Another example is Golem, a network for renting out spare computing power to others, with its own token called Golem Network Token (GNT), which is used for payments on this network.

---

[10] The term ""Big Data" mall" was used in [Informatica, 2011]. However, there it was used in the sense of a mall for Big Data in a marketplace (which is not a data marketplace) that offers other (non-data) products and services.
[11] This is not to say that all data must be in the same format – this too would be impracticable as data types vary.

Much in the same spirit, Coin Economies can also be useful for decentralized Data Malls. The payments on a decentralized Data Mall network are made through its coin[12] issued, say, via Ethereum. If this coin is exchange-traded, then a data vendor can convert tokens received from data sales into fiat currency, so in this case there is no issue with liquidity. Instead, there can be an issue with volatility – if the exchange-traded coin is volatile, the vendor can end up receiving less if the coin value drops (or more, if the coin value rises). There are ways to mitigate this via, e.g., employing derivatives hedging strategies such as buying put options on the coin (if such options are available for it), which is known as the "protective put" or "married put" strategy (see, e.g., [Cohen, 2005], [Figlewski, Chidambaran and Kaplan, 1993]). Alternatively, the vendor can hedge by selling futures (exchange traded) or a forward contract (over-the-counter) (see, e.g., [Hull, 2012]), if there is a market for such derivatives, thereby forgoing the upswing upside.

If the coin is not exchange-traded, then liquidity becomes an issue. After all, the vendor wishes to monetize its data, not just acquire illiquid tokens. There is an old-fashioned simple solution for this, akin the good ol' Chuck E. Cheese tokens. Suppose a vendor collected $S$ dollars worth of tokens through sales. With proof of sale, the token issuer will swap these tokens for $P$ dollars in fiat money,[13] where (also see Figure 2)

$$P = S \times (1 - C) \tag{1}$$

Here $C$ is some reasonable predefined percentage. In essence, $C$ resembles a sales commission percentage and the token issuer makes $Q$ dollars from the vendor's sale, where

$$Q = S \times C \tag{2}$$

Except that $Q$ is paid not for sales, but for access to the network: the vendor "outsources" sales to the network, it is not required to hire salespeople, etc. The network creator gets rewarded for creating the network, just, e.g., as with Auger or Golem. Let us mention that Eqns. (1), (2) do not capture all transaction fees associated with the sale. E.g., with Ethereum-based tokens each user needs a wallet to transact in such a token, and transaction fees are paid to the Ethereum blockchain miner of the block containing the user's smart contract. These transaction fees are paid in ETH [Ethereum, 2018] and used to be small, even recently [Young, 2017]. However, the cryptocurrency prices have skyrocketed lately, and so have the fees, including for ETH (see, e.g., [BitInfoCharts, 2018]). One issue with rapidly rising absolute (as opposed to relative to the ETH price) transaction fees is their adverse effects on the viability of micropayments using ETH.

---

[12] By "coin" we mean coin or token and vice-versa, as may be applicable best.
[13] In practice, with, e.g., Ethereum-based tokens, this payment can be made in ether (ETH) and then the vendor can convert ETH to fiat money. Both the vendor and network creator have exchange rate exposure (see above).

Increasing transaction fees, so long as they remain within reason, albeit unpleasant, do not affect big-ticket data sales (e.g., a datafeed that costs $100,000/yr). But they can adversely affect small deals requiring micropayments. This includes pay-per-use cases such as access to blogs, research reports, etc., which can be valuable to a number of users provided that they are reasonably priced (with monetization through volume rather than high margins). Being able to make micropayments without incurring prohibitive transaction costs is paramount to such uses.

Are networks such as Ethereum going to adjust to the changing reality (cryptocurrency prices skyrocketing) and change the transaction fee structure? It is unclear how realistic this is considering that micropayments are not the biggest sources of revenues for the miners. There are efforts to fill this niche, e.g., μRaiden (MicroRaiden), which specializes in micropayments on the Ethereum network [Silver, 2017]. Unsurprisingly, this is an "off-blockchain" solution. In this regard, a Data Mall network creator may as well consider implementing its own solution. In fact, one may further argue that the entire blockchain solution can be implemented organically as opposed to using a network such as Ethereum. We can take this line of reasoning further and consider an organic blockchain solution with no mining: all transactions are verified by a central authority (the network creator), thereby completely eliminating transaction fees in P2P or B2B (business-to-business) payments.[14] Micropayments are then no less viable than large payments. The network creator's efforts are compensated via the fees collected from vendors (Eqn. (2)).

Would this not defy the purpose of blockchain – decentralization – in the first instance? This is not an "ideological" question but a pragmatic one.[15] To the network creator the appeal of the blockchain technology is mainly in that it does not have to reinvent the wheel and can use an open source solution as opposed to building its own ledger from scratch. For the users (both the vendors and the data consumers) it is not important that the transactions are mined – so long as the accounting is done right, it does not really matter whether it is done by miners or in a centralized fashion. In the context of a Data Mall, the users simply want a solution with lower costs. Blockchain can still play a sizable role from the users' perspective (as opposed to that of the network creator) in this essentially "hybrid" approach: i) it provides a backbone for a decentralized Data Mall; and ii) it can record all transactions for transparency (without mining).

---

[14] One solution is to have 2 tokens: Ethereum-based "class A" for liquidity, and internal "class B" for transactions. The "class B" token need not be blockchain-based. It can be maintained in any database by the central authority.
[15] In this regard, note that many cryptocurrencies (or, more precisely, coins/tokens) are not mined (see Table 1).

We emphasize that the "hybrid" approach we discuss above is not confined only to Data Malls but can have much broader applicability to other types of networks. Also, blockchain is not the only available approach and others have both been proposed and implemented. Ripple, for instance, is not based on blockchain; instead, it is a consensus based payment protocol (see, e.g., [Armknecht *et al*, 2015]). We already mentioned the IOTA Foundation's Tangle technology above. There are other alternative technologies as well: Hashgraph (more centralized but also reportedly more scalable than blockchain) [Baird, 2016], the Ceptr project [P2P Foundation, 2017], etc. A pragmatic approach to a given problem should dictate a solution, not "ideology".

## 4. Data Provenance

Data provenance/validation is one of the contemplated applications of blockchain (see, e.g., [Azaria *et al*, 2016], [Dooley, 2017], [ECS, 2017], [Gipp, Jagrut and Breitinger, 2016], [Kuo, Kim and Ohno-Machado, 2017], [Lemieux, 2016], [Liang *et al*, 2017], [Neisse, Steri and Nai-Fovino, 2017], [Ramachandran and Kantarcioglu, 2017], [Rogers, 2015], [Wilkinson and Lowry, 2014], [Zikratov *et al*, 2017], [Zyskind, Nathan and Pentland, 2015]). In this context, not all data can be treated on the same footing. Generally, when discussing large datasets, as we already alluded to above, it would be incorrect to assume that such datasets are stored in blockchain as this is simply impracticable. Furthermore, a P2P (or even B2B) network is simply not suited for validating many types of datasets in the first instance. Consider financial data, e.g., intraday tick data from NYSE (New York Stock Exchange). This dataset is large. Furthermore, this data is not something an average user can validate. There can be issues with this dataset as provided by a vendor or the exchange itself as errors can and do occur, without any malice or ill-intention on any actor's part. Depending on the type of error, it may or may not be corrected, and such a correction usually occurs at the institutional level. More generally, data vendors retroactively do correct errors in datasets and this is a well-known issue in, e.g., quantitative trading, where such retroactive corrections introduce "in-sampleness" into historical data. I.e., a quantitative researcher attempting to backtest a strategy may be working with different data than what would be available at the time the trading took place historically, and this can introduce biases into the backtest that are not even measurable. Usually such biases are small and one just lives with them. *Así es la vida*. It would be utterly impractical to put such a process on blockchain.

However, there are cases where blockchain could play a role in data provenance. In this regard, it is important to take a pragmatic approach and accept as a fact of life that populating blockchain with large data is impracticable. What is practicable is to place some much smaller metadata on blockchain. This metadata can, e.g., describe the changes made to the data. Wiki databases are viable examples of this application. Such databases can contain large amounts of data and it is impracticable to put it all on blockchain. Nor is there any pressing need to do so from a pragmatic viewpoint. However, recording the history/metadata associated with edits to such a wiki database can very well be practicable (assuming the recorded data is reasonably concise). E.g., the history of Wikipedia entry edits could possibly be recorded on blockchain. In fact, Everipedia is planning to convert to EOS blockchain (which will store not all information, but lighter data) while using the Interplanetary File System (IPFS) (for storing data-heavy files such as images and videos) [Del Castillo, 2017b], [Hertig, 2017], [Rubin, 2017], [Stanley, 2017].

This kind of approach, which in principle is practicable, can be applied to other types of networks, including Data Malls. Whether using IPFS is required would depend on a given situation: notwithstanding the hype, there is nothing intrinsically wrong with private entities storing their data on secure servers, properly backed up, with failover protocols in place, etc. Not everything needs to be decentralized. Pragmatic considerations should be the guide in this.

Everipedia, which is Wikipedia's main competitor, will also use its own token called "IQ" for payments on its network. IQ will be used to incentivize content creation. IQ is going to be minable: it will be mined by editors, curators, etc., by making accurate, valuable contributions to "the encyclopedia of everything", as Everipedia refers to itself. This too is a Coin Economy.

## 5. Keyless Payments

With a proliferation of cryptocurrencies and tokens, things got a bit "messier" than before. If one wishes to transfer, say, BTC (or any other cryptocurrency or token) to another person or entity, one needs to know the recipient's virtually-impossible-to-remember BTC address. This very issue – conceptually, that is – existed well before cryptocurrencies: bank account numbers are also not something people carry in their memory. PayPal has a simple solution: a PayPal account is linked to an email address, bank account numbers are stored internally, so sending money to any user with a PayPal account requires knowing only the recipient's email address.

The Venmo mobile app (Venmo was first acquired by Braintree in 2012 with PayPal acquiring Braintree in 2013 [Gillette, 2014]) has a model similar to PayPal. Various big banks internally provide a similar service using clearXchange/Zelle app technology [Robin, 2016]. Users who have Apple Pay set up can send money to each other using Apple's iMessage [Chowdhry, 2017]. One can use Facebook Messenger to send money via PayPal [Gajanan, 2017]. And so on. Conceptually, what all these services do is that they allow users to associate an email addresses or some other easy-to-remember IDs/handles with their digital wallets that store all the other required banking information (account numbers, routing numbers, etc.).

With cryptocurrencies and tokens one also deals with digital wallets. However, there is no service similar to PayPal or others described above where all one needs is the recipient's email address or some other easy-to-remember ID/handle.[16] A mobile app providing such a service would be very convenient. Just as with, say, PayPal, users would create digital wallets, which can include their various cryptocurrency addresses (and can also include bank accounts). Each wallet would have what we refer to as a Keyless Handle – an email address and/or some other easy-to-remember ID/handle – associated with it. Sending cryptocurrency (or fiat money) would require only knowing the recipient's Keyless Handle. This is what we refer to as Keyless Payments. In this day and age, while a web-based service might be a plus, a mobile app would be required for Keyless Payments. This app can, e.g., be a multimedia messenger. Sending BTC,

---

[16] There are web-based services for sending BTC "by email", etc., e.g., MoneyBadger.io. The sender provides the recipient's email and sends BTC to a BTC address provided by MoneyBadger, then MoneyBadger sends an email with a code to the recipient, the recipient enters the code along with a cryptocurrency address (BTC, ETH, etc.) on MoneyBadger's website, and MoneyBadger transfers the corresponding cryptocurrency to the recipient's address.

ETH, other cryptocurrencies, tokens and fiat money would be as easy as sending a message.[17] Such a messenger may or may not be blockchain-based. In certain contexts a blockchain-based messenger can be useful, e.g., if messages must be indelibly logged for compliance purposes.

Let us emphasize that any centralized conversion service such as Keyless Payments can be prone to being hacked. However, conceptually, this is no different from blockchain actually not being an infallible solution when it comes to practical applications. Thus, while blockchain per se may be impractical to hack, e.g., exchanges where blockchain-based cryptocurrencies are traded can be and have been hacked. Mt. Gox's infamous $460M disaster [McMillan, 2014] is only one example of cryptocurrency exchanges being hacked. The same applies to any digital wallet, whether it contains cryptographic keys, bank accounts or other sensitive information. A foolproof way of storing a cryptoasset is to deliberately forget its key. It is also utterly useless. In this regard, at least conceptually, Keyless Payments would be no different from PayPal, etc.[18]

## 6. (Proxy) Voting

A desire to apply blockchain to voting is natural considering its fraud-resistant nature. Such applications have been discussed in, e.g., [Barnes, Brake and Perry, 2016], [Ben Ayed, 2017], [Boucher, 2016], [Firth, 2017], [Gabison, 2016], [Higgins, 2017], [Koven, 2016], [Lee *et al*, 2016]. Here we will not delve into voting in the political context. Instead, we focus on corporate (proxy) voting. It is no secret that small shareholders are effectively disenfranchised for all intents and purposes (see, e.g., [Kakushadze, 2015], [Mourning, 2007], [Wilcox, 2004], [Wink and O'Leary, 2009]) as they either do not actually vote or, if they do, it is done though a proxy vote (see, e.g., [Investopedia, 2018]). This archaic process, especially when a proxy ballot and a proxy statement are sent by mail,[19] is cumbersome and many small shareholders do not even bother. Large shareholders – the movers and shakers – then control the company without input from many smaller shareholders, who effectively are disenfranchised, even if unintentionally.

---

[17] For some literature on mobile payments, see, e.g., [Au, 2008], [Carton *et al*, 2012], [Chandra, Srivastava and Theng, 2010], [Chen, 2008], [Crowe, Rysman and Stavins, 2010], [Dahlberg *et al*, 2008], [De Albuquerque, Diniz and Cernev, 2014], [De Reuver *et al*, 2015], [Dennehy and Sammon, 2015], [Dewan and Chen, 2005], [Donner and Tellez, 2008], [Gowal and Gowal, 2014], [Hayashi, 2012], [Isaac, and Zeadally, 2014], [Jacob, 2007], [Jaradat and Al-Mashaqba, 2014], [Karnouskos, 2004], [Karnouskos *et al*, 2008], [Kemp, 2013], [Kim, Mirusmonov and Lee, 2010], [Kreyer, Turowski and Pousttchi, 2003], [Laukkannen, 2008], [Liébana-Cabanillas, Sánchez-Fernández and Muñoz-Leiva, 2014], [Liu, 2016], [Liu, Kauffman and Ma, 2015], [Lu *et al*, 2011], [Mallat, 2007], [Mallat and Tuunainen, 2008], [Mbogo, 2010], [Niranjanamurthy, 2014], [Ondrus and Pigneur, 2006], [Polasik *et al*, 2013], [Pousttchi, 2008], [Rodrigues *et al*, 2014], [Schierz, Schilke and Wirtz, 2010], [Shaw, 2014], [Shin, 2009], [Slade *et al*, 2015], [Tan *et al*, 2014], [Thakur and Srivastava, 2014]. See [Hanly, 2017] on Square accepting mobile payments in BTC.
[18] Albeit one's bank account enjoys robust protections from fraudulent activities, which cryptoassets do not afford.
[19] Not to mention all the trees this kills.

A foolproof online voting system would allow shareholders to be more engaged. This is where blockchain technology can be useful. There appears to be concrete progress made in this direction; see, e.g., [AST, 2017], [Crichton, 2017], [De, 2017a], [Kovlyagina and Yakovlev, 2016], [NASDAQ, 2017], [Rizzo, 2017]. This is an interesting application of blockchain technology. What remains to be seen is if it sticks. It is important to consider that without mining "many in-house blockchain solutions will be nothing more than cumbersome databases" [Hampton, 2016].[20]

## 7. Concluding Remarks

In this paper we discussed some applications and aspects of blockchain. Other applications have been contemplated. Table 2 summarizes a partial list of areas where blockchain has been or may be implemented. Some of these applications will probably withstand the test of time, some may not. What is clear is that blockchain technology has already made a huge impact – even though some of this impact could be very transient – and is likely to continue to do so for years to come as it appears to appeal to and excite scores of enthusiasts, entrepreneurs, mom-and-pop as well as professional investors, technologists, scholars, futurists, etc. Perhaps one of the reasons is captured in by-now-"iconic" schematic depiction of "centralized vs. decentralized vs. distributed" (see Figure 3). Let us conclude with our message: "Pragmatism over Ideology!"

---

[20] Also see, e.g., [Stinchcombe, 2017].

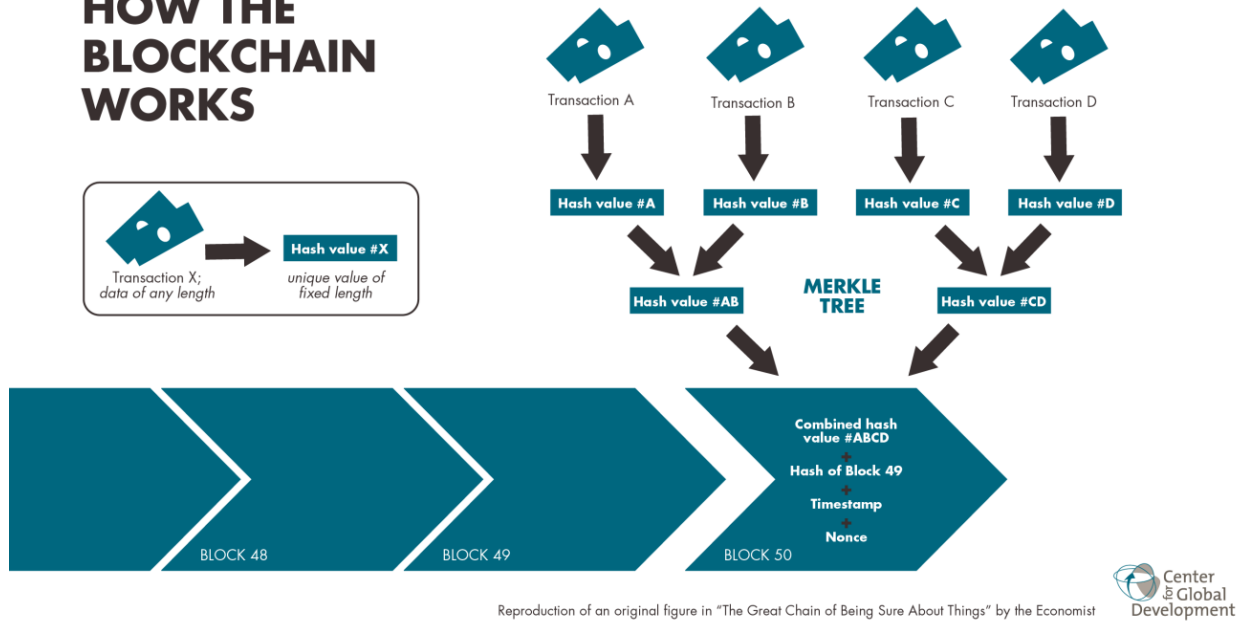

**Figure 1.** A schematic depiction of blockchain. Attribution: By B140970324 (Own work) [CC BY-SA 4.0 (https://creativecommons.org/licenses/by-sa/4.0)], via Wikimedia Commons. Image source: https://upload.wikimedia.org/wikipedia/commons/3/3a/Blockchain_workflow.png.

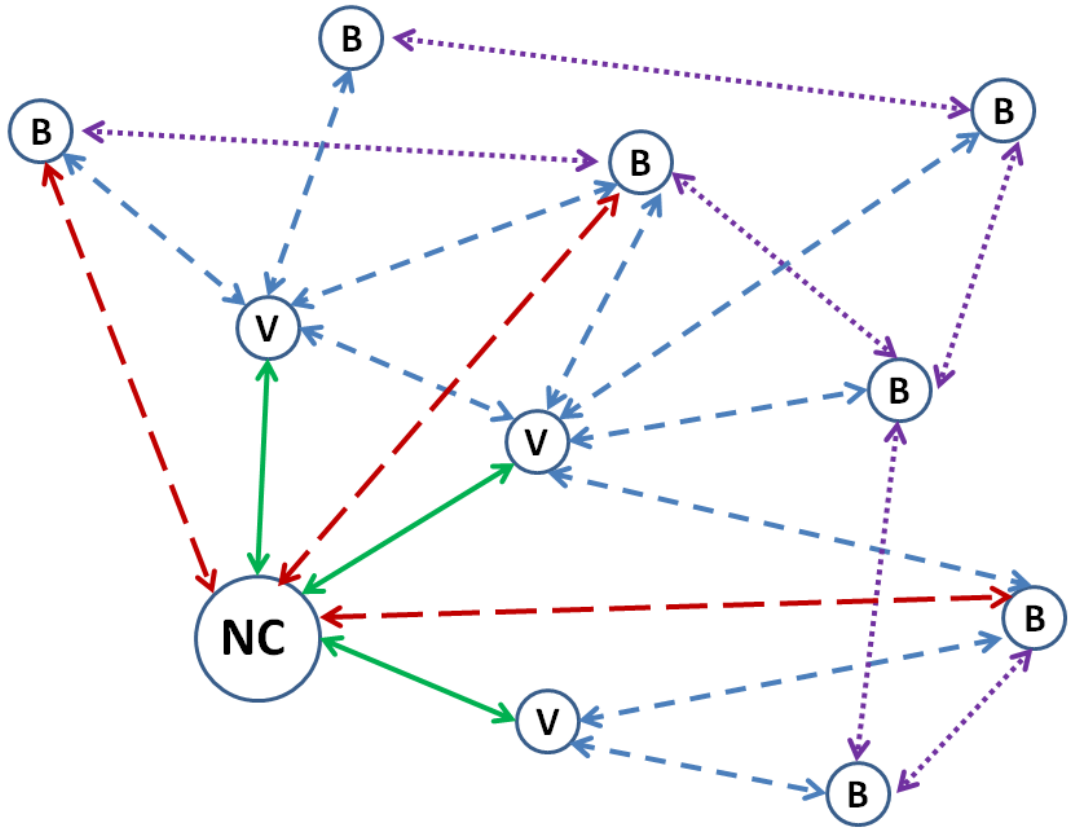

**Figure 2.** A schematic depiction of a Data Mall network. NC = network creator; V = data vendor; B = data buyer. Solid lines represent token payments from vendors to the network creator and fiat money (or exchange-traded coin) payments from the network creator to the vendors (see Section 3). Smaller dashed lines represent token payments from buyers from vendors and data provision by vendors to buyers. Larger dashed lines represent miscellaneous interactions between buyers and the network creator. Dotted lines represent miscellaneous interactions between buyers. Each buyer may acquire data from more than one vendor. A vendor can also be a buyer and acquire data from another vendor. And so on.

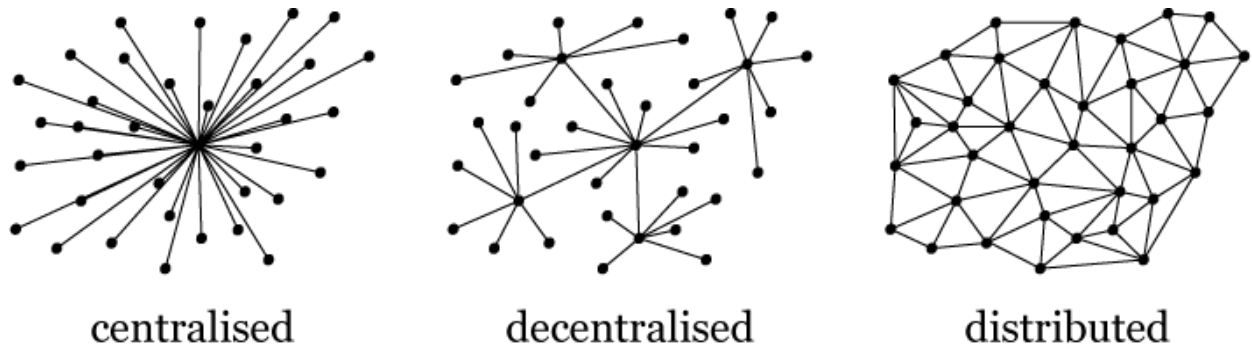

**Figure 3.** A schematic depiction of "centralized vs. decentralized vs. distributed". Attribution: 1983~enwiki at English Wikipedia [GFDL (http://www.gnu.org/copyleft/fdl.html), CC-BY-SA-3.0 (http://creativecommons.org/licenses/by-sa/3.0/) or CC BY-SA 2.5-2.0-1.0 (https://creativecommons.org/licenses/by-sa/2.5-2.0-1.0)], via Wikimedia Commons. Image source: https://upload.wikimedia.org/wikipedia/commons/b/ba/Centralised-decentralised-distributed.png.

| Rank | Name | Symbol | Market Cap, $B | Price, $ | Volume (24h), $M | Mined |
|------|------|--------|----------------|----------|-------------------|-------|
| 1 | Bitcoin | BTC | 289.156 | 17,226.10 | 20,295 | Yes |
| 2 | Ripple | XRP | 119.344 | 3.08 | 4,135 | No |
| 3 | Ethereum | ETH | 101.746 | 1,050.98 | 5,269 | Yes |
| 4 | Bitcoin Cash | BCH | 47.069 | 2,785.68 | 1,635 | Yes |
| 5 | Cardano | ADA | 26.602 | 1.03 | 365 | No |
| 6 | Litecoin | LTC | 16.590 | 303.61 | 2,374 | Yes |
| 7 | NEM | XEM | 14.663 | 1.63 | 119 | No |
| 8 | Stellar | XLM | 12.950 | 0.724340 | 595 | No |
| 9 | TRON | TRX | 11.414 | 0.173607 | 2,675 | No |
| 10 | IOTA | MIOTA | 10.993 | 3.96 | 181 | No |

**Table 1.** Top 10 cryptocurrencies by market capitalization. The data (market capitalization and volume are rounded) is taken from [CoinMarketCap, 2018] as of approximately 11:51 PM EST on January 6, 2018. Note that cryptocurrency prices are volatile and this data is a snapshot.

| Use | Some references (also see references therein) |
|-----|-----------------------------------------------|
| Decentralized cryptocurrencies | See Section 1, [Andreesen, 2014], [Krugman, 2013], [Vigna and Casey, 2015], [Wang and Vergne, 2017] |
| Government-issued cryptocurrencies | See Section 1 |
| Financial transactions (other than cryptocurrencies) | [De, 2017b], [Del Castillo, 2017a], [Martin, 2016], [Rohlfing and Davis, 2017], [Ito, Narula and Ali, 2017] |
| Supply chains | [Crosby *et al*, 2016], [Nash, 2016], [Tian, 2017] |
| Social networking | [Olenski, 2017], [Tapscott and Tapscott, 2016] |
| Business processes/management | [Beck *et al*, 2017], [Guo and Liang, 2016], [Mendling *et al*, 2017], [Prybila *et al*, 2017], [Weber *et al*, 2016] |
| Governance | [Atzori, 2017], [Piazza, 2017], [Yermack, 2017] |
| Prediction markets | See Section 3 |
| Voting | See Section 6 |
| Proxy voting | See Section 6 |
| Medical records | [Al Omar *et al*, 2017], [Azaria *et al*, 2016], [Hoy, 2017], [Orcutt, 2017], [Xia *et al*, 2017], [Yue *et al*, 2016] |
| Smart contracts | See Section 3 |
| Data Malls | See Sections 2 and 3 |
| Data marketplaces (IoT) | See Section 2 |
| Shared computing | See Section 3 |
| Company incorporation/registration | [Roberts, 2017], [Ryan, 2015], [Vigna, 2016] |
| Copyrighted works distribution | [MBW, 2017], [Willms, 2016], [Zellinger, 2016] |
| Data storage/transfer | [Ali *et al*, 2017], [Callahan, 2017], [Romano and Schmid, 2017], [Sengupta *et al*, 2016], [Suberg, 2017] |
| Data provenance | See Section 4 |
| Decentralized Autonomous Organizations | [Chohan, 2015], [Deegan, 2014], [DuPont, 2018], [Harrison, 2016], [Jentzsch, 2016], [Merkle, 2016] |

**Table 2.** A partial list of the areas with (potential) blockchain uses.